\documentstyle[multicol,prl,aps,epsfig,subfigure]{revtex}
  
\begin{document}
\bibliographystyle{unsrt} 



\title{
\bf \Large Determination of the Deep Inelastic Contribution to the Generalised
Gerasimov-Drell-Hearn Integral for the Proton and Neutron
}

\maketitle
\begin{center}
\centerline{\it The HERMES Collaboration}
K.~Ackerstaff$^{5}$, 
A.~Airapetian$^{34}$, 
N.~Akopov$^{34}$, 
I.~Akushevich$^6$,
M.~Amarian$^{25,29,34}$, 
E.C.~Aschenauer$^{6,13,25}$, 
H.~Avakian$^{10}$, 
R.~Avakian$^{34}$, 
A.~Avetissian$^{34}$, 
B.~Bains$^{15}$
% ...
\\
\small{\em
% ...
$^5$DESY, Deutsches Elektronen Synchrotron, 22603 Hamburg, Germany\\
$^6$DESY Zeuthen, 15738 Zeuthen, Germany\\
% ...
$^{10}$Istituto Nazionale di Fisica Nucleare, Laboratori Nazionali di Frascati, 00044 Frascati, Italy\\
% ...
$^{13}$Department of Subatomic and Radiation Physics, University of Gent, 9000 Gent, Belgium\\
% ...
$^{15}$Department of Physics, University of Illinois, Urbana, Illinois 61801, USA\\
% ...
$^{25}$Nationaal Instituut voor Kernfysica en Hoge-Energiefysica (NIKHEF), 
1009 DB Amsterdam, The Netherlands\\
% ...
$^{29}$Istituto Nazionale di Fisica Nucleare, Sezione Sanit\`a, and Istituto Superiore di Sanit\`a, Physics Laboratory, 00161 Roma, Italy\\
% ...
$^{34}$Yerevan Physics Institute, 375036, Yerevan, Armenia\\
}
\end{center}

\begin{abstract} 
The virtual photon absorption cross section differences 
$[\sigma_{1/2}-\sigma_{3/2}]$ for the proton and neutron 
have been determined from measurements of polarised cross section asymmetries
in deep inelastic scattering of 27.5 GeV longitudinally 
polarised positrons from polarised $^1$H and $^3$He internal gas targets.
The data were collected in the region above the nucleon resonances 
in the kinematic range $\nu <$ 23.5 GeV and 0.8 GeV$^2 < Q^2 <$ 12 GeV$^2$.
For the proton the contribution to the generalised Gerasimov-Drell-Hearn 
integral was found to be substantial and must be included for an accurate 
determination of the full integral.
Furthermore the data are consistent with a QCD next-to-leading order fit
based on previous deep inelastic scattering data.
Therefore higher twist effects do not appear significant.

\end{abstract}


\begin{multicols}{2}[]

The GDH sum rule, derived in the sixties by Gerasimov \cite{Geras} and
independently by Drell and Hearn \cite{DreHea}, relates the anomalous 
contribution $\kappa$ to the magnetic moment of the nucleon 
($\kappa_{\mathrm p}$=1.79, $\kappa_{\mathrm n}$=$-$1.91)
with the total absorption cross sections for circularly polarised real
photons on polarised nucleons.
It is written as:

\begin{equation}
\label{gdhorig}
\int_0^{\infty}
\Bigl[\sigma_{1/2}(\nu)-\sigma_{3/2}(\nu)\Bigr]\frac{d\nu}{\nu}=
-\frac{2\pi^2\alpha}{M^2}\kappa^2,
\end{equation}
where  $\sigma_{1/2}$ and $\sigma_{3/2}$
are the  photo-absorption cross sections for total helicities 1/2 and 3/2, 
$\nu$ is the photon energy, and $M$ is the nucleon mass. 
The sum rule arises from the combination of a general 
dispersion relation for forward Compton scattering \cite{GGT} 
and the low-energy theorem of Low \cite{LOW}, with the additional 
assumption that the spin-flip part of the Compton  scattering amplitude goes
to zero at infinite energy without any poles.
The theoretical predictions for the integral are --204 $\mu$b and --233 $\mu$b 
for the proton and neutron, respectively.
Experimentally this sum rule has never been tested directly because of the 
lack of suitable polarised real photon beams. 

The integral defined in Eq.~\ref{gdhorig} can be generalised to the absorption 
of virtual photons with energy $\nu$ and squared four-momentum $-Q^2$:

\begin{eqnarray}
\label{gdhgen}
I(Q^2) \equiv \int_{Q^2/2M}^{\infty}
\Bigl[\sigma_{1/2}(\nu,Q^2)-\sigma_{3/2}(\nu,Q^2)\Bigr]\frac{d\nu}{\nu}
\nonumber 
\\
 = \frac{8 \pi^2 \alpha}{M} \int_0^1 \frac{g_1(x,Q^2) - 
\gamma^2 g_2(x,Q^2)}{K} \frac{dx}{x}.
\end{eqnarray}
Here $g_1$ and $g_2$ are the polarised structure functions of the nucleon,
$x$ is the Bjorken variable and $K = \nu \sqrt{1 + \gamma^2}$ is the flux 
factor of the virtual photons \cite{GIL}, with $\gamma^2=Q^2/\nu^2$.
For $Q^2$=0 the left side of Eq.~\ref{gdhgen} is trivially equal to the left 
side of Eq.~\ref{gdhorig}.
The $Q^2$ dependence of this integral connects the static ground state 
properties of the nucleon with its helicity structure as measured in inelastic
scattering in the resonance and deep inelastic regions.

Interest in the generalised GDH integral arose in 1989 when it was related
\cite{ANSELM} to $\Gamma_1 = \int_0^1 g_1(x)dx$, which had been 
measured \cite{EMC,E130} to be significantly smaller than expected 
\cite{EJ}.
In fact, for $\gamma \ll$ 1, the right side of Eq.~\ref{gdhgen} reduces to 
$I_1(Q^2) \equiv 16 \pi^2 \alpha \Gamma_1 / Q^2$.
For both neutron and proton a strong variation of $I(Q^2)$ is required in 
order to connect $\Gamma_1$ to the GDH prediction for real photons; in the case
of the proton $\Gamma_1$ is positive and then $I(Q^2)$ must change sign
at low $Q^2$.
Several possible explanations of this behaviour have been proposed [10-16] 
-- for example the contributions of the resonances, of $g_2$ and of higher 
twist effects to the $Q^2$-evolution of the integral were taken into 
consideration.

This letter presents results for the cross section differences
$[\sigma_{1/2}-\sigma_{3/2}]$ extracted from measurements of the longitudinal
positron-nucleon cross section asymmetries $A_{\parallel}$ 
for the neutron and the proton in the deep inelastic scattering (DIS) region.
The data were collected during the 1995 and 1997 years of operation
of the HERMES experiment.
The 1995 $^3$He data for the neutron have previously been used for the 
extraction of the polarised structure function  
$g_1^{\mathrm n}(x)$ \cite{HERMES}; the 1997 $^1$H data have been used for the 
determination of $g_1^{\mathrm p}(x)$ \cite{A1p}.

The experiment was performed with a 27.5 GeV beam of longitudinally 
polarised positrons incident on longitudinally polarised $^1$H and $^3$He gas 
targets internal to the HERA storage ring at DESY.
The positron beam in the HERA ring is transversely polarised by emission of
synchrotron radiation \cite{ST}. 
Longitudinal polarisation is obtained by using spin rotators located upstream 
and downstream of the HERMES experiment \cite{BA}. 
The average beam polarisation for the analysed data was 0.55 with a relative
systematic uncertainty of 3.4\% (3.9\%) for the proton (neutron).

The $^1$H target atoms are produced by an atomic beam source (ABS) based on 
Stern-Gerlach separation of atomic hydrogen spin states \cite{STO}.
The $^3$He target atoms are polarised by spin exchange collisions with
optically-pumped $^3$He atoms in the 2$^3$S meta-stable state \cite{xx97}.
The beam of polarised atoms enters a 400 mm long open-ended thin-walled 
storage cell located inside the storage ring,
providing an areal target density of approximately 
$7 \times 10^{13}$ $^1$H-atoms/cm$^2$ or
$3.3 \times 10^{14}$ $^3$He-atoms/cm$^2$. 
The average values of the target polarisation during the experiment were 
0.88 $\pm$ 0.04 for $^1$H \cite{A1p} and 0.46 $\pm$ 0.02 for $^3$He 
\cite{HERMES}.

The magnetic spectrometer is fully described in Ref.~\cite{yy97}.
It is constructed as two identical halves mounted above and below the 
positron ring plane.
The angular acceptance of the spectrometer extends over the range
40 mrad $< \theta < $ 220 mrad. 
For positrons the average angular resolution is better than 1 mrad and the
average momentum resolution is better than 2\% aside from bremsstrahlung
tails.
Positron identification is accomplished with a lead glass  
calorimeter, a preshower counter, a transition radiation detector 
and a gas threshold \v{C}erenkov counter. 
For the analysed data sets the hadron contamination was less than 1\% at an
average positron identification efficiency of 99\%.

The cross section difference $[\sigma_{1/2}-\sigma_{3/2}]$ can be expressed in 
terms of the virtual photo-absorption asymmetry $A_1$ and the unpolarised 
structure function $F_1$:
\begin{equation}
\label{dsigma}
\sigma_{1/2}-\sigma_{3/2} = \frac{8\pi^2\alpha}{M}\;\frac{A_1F_1}{K}.
\end{equation}
The structure function $F_1 = F_2(1+\gamma^2)/(2x(1+R))$ was 
calculated from published parameterisations of the unpolarised structure 
function $F_2$ \cite{NMCF2} and of $R = \sigma_L/\sigma_T$ \cite{WHITL}, 
the ratio of the absorption cross sections for longitudinally and transversely 
polarised virtual photons. 
The asymmetry $A_1$ was extracted from the measured longitudinal asymmetry 
$A_{\parallel}$ by means of the formula $A_1 = A_{\parallel}/D - \eta A_2$,
where 
$ D = y(2-y)(1+\gamma^2y/2)/[y^2(1+\gamma^2)(1-2m_e^2/Q^2)+
      2(1-y-\gamma^2y^2/4)(1+R)]$ 
is the virtual photon depolarisation factor ($m_e$ is the electron mass),
$\eta = \gamma(1-y-\gamma^2y^2/4)/(1-y/2)(1+\gamma^2y/2)$ 
and $y=\nu/E$, where $E$ is the beam energy.
The asymmetry $A_2$ is bounded by the positivity limit $A_2\leq \sqrt{R}$ and 
has been measured to be small [26-28]; moreover, the factor $\eta$
is less than 0.5 in the kinematic range covered by this analysis. 
The contribution of $A_2$ for the proton was evaluated with a parameterization
$A_2^{\mathrm p} = 0.5x/\sqrt{Q^2}$ based on existing data \cite{143,SMC},  
while the contribution of $A_2$ for the neutron was neglected 
and the uncertainty \cite{A2} was included in the systematic error.

\begin{figure}[htb]
\begin{center}
\epsfxsize 8 cm {\epsfbox{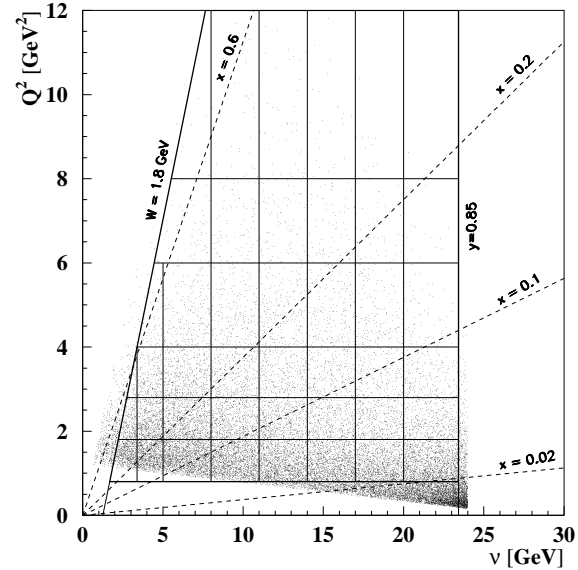}}
\begin{minipage}[r]{\linewidth}
\caption{Kinematic plane covered  by the HERMES experiment showing the
boundaries and the binning in $Q^2$ and $\nu$ used for the evaluation of the 
GDH integral for the proton data above the resonance region.
The points shown are a sample of the whole set of collected data.}
\label{nubins}
\end{minipage}
\end{center}
\end{figure}

After applying data quality criteria, 1.8 (2.7) million events for the proton 
(helium) target were available in the kinematic range 
0.8 GeV$^2$ $< Q^2 <$ 12 (10) GeV$^2$, $\nu <$ 23.5 GeV and
$W > W_0 = $ 1.8 (2) GeV, where $W$ is the energy of the hadronic final 
state.
The kinematic plane covered by this experiment is shown in Fig.~\ref{nubins}. 
Above the minimum angle of 40 mrad the events are continously distributed 
allowing -- without interpolation -- the integration over $\nu$ for 
several bins in $Q^2$.

The $x$-range covered in the different $Q^2$-bins varied from 0.03$-$0.36 
for the lowest $Q^2$-bin to 0.2$-$0.8 for the highest one.
The kinematic range for the proton was slightly increased compared to that of
the neutron due to the intrinsically higher figure of merit of the $^1$H
target.

The value of $A_{\parallel}/D$ has been extracted in each bin by using

\begin{equation}
\label{eq:expA1}
\frac{A_{\parallel}}{D}  = \frac{1}{D}
\frac{N^{-}L^{+}-N^{+}L^{-}}{N^{-}L_P^{+}+N^{+}L_P^{-}},
\end{equation}
where $N$ is the number of detected scattered positrons corrected for 
$e^+e^-$ background from charge symmetric processes.
Here $L$ is the integrated luminosity corrected for dead time; $L_P$ is the 
integrated luminosity corrected for dead time and weighted by the product of 
beam and target polarisations. 
The superscript $+$ $(-)$ refers to the situation where the target spin axis 
was oriented parallel (anti-parallel) to that of the positron beam.
Radiative corrections are typically 2\% of the observed asymmetry for the 
proton and 20\% for the neutron. They were calculated using the prescription
given in Ref. \cite{POLRAD}. 
The neutron asymmetry $A_1^{\mathrm n}$ was obtained from the $^3$He asymmetry
by correcting for nuclear effects, assuming a relative polarisation of 
0.86 $\pm$ 0.02  for the neutron and $-$0.028 $\pm$ 0.004 for each of the two 
protons in the $^3$He nucleus \cite{FRIAR}, and using a fit of data for 
$A_1^{\mathrm p}$ \cite{E143Q2}.

\begin{figure}[hb]
\begin{center}
\epsfxsize 9.0 cm {\epsfbox{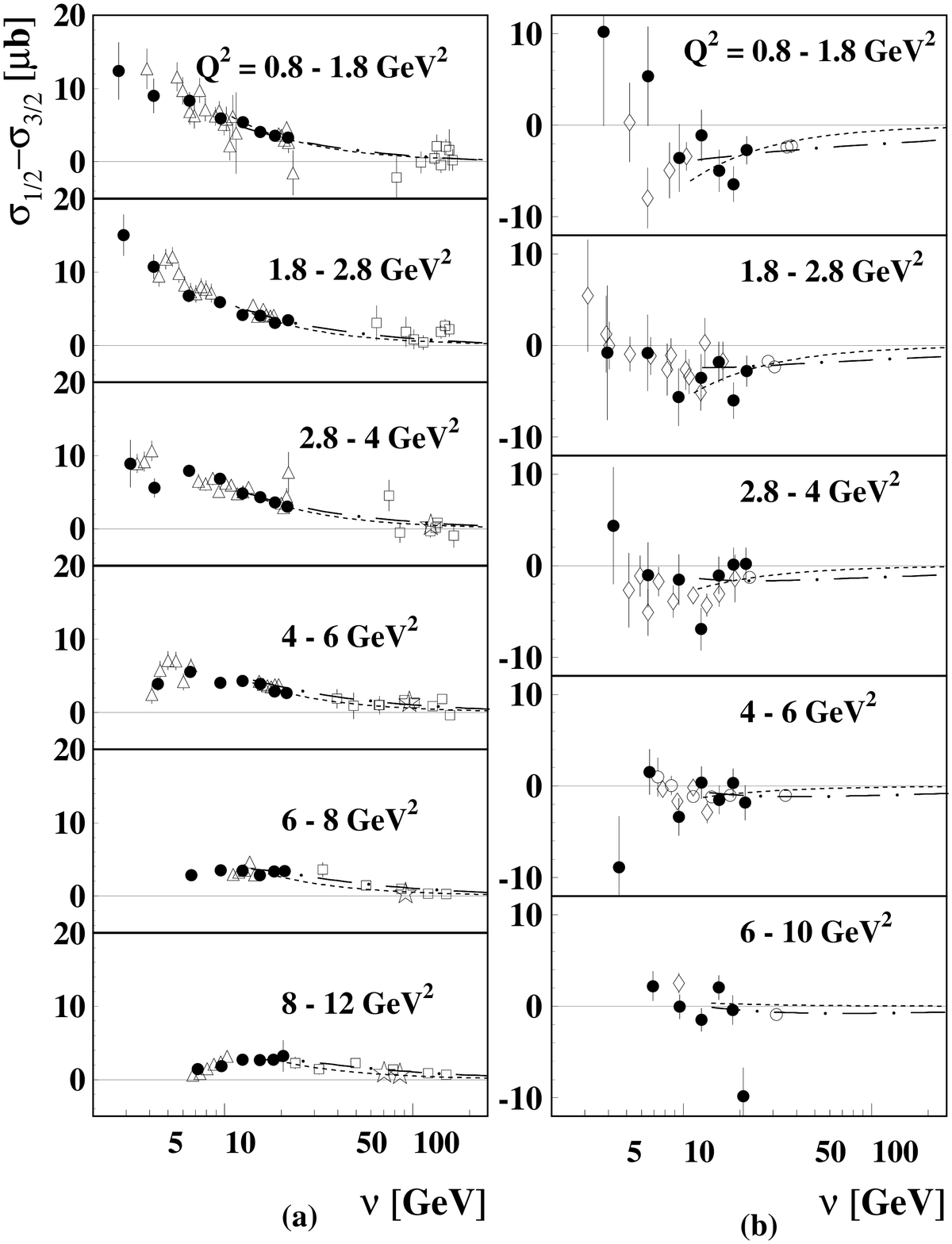}}
\begin{minipage}[r]{\linewidth}
\caption{Cross section differences as a function of $\nu$ measured
         in different bins of $Q^2$ for the proton (a)
         and the neutron (b). 
         Filled circles are data from this experiment.
         Open symbols are values derived from other experiments:
         stars \protect\cite{EMC},  
         triangles \protect\cite{143}, squares \protect\cite{SMC}, 
         diamonds \protect\cite{An}, and circles \protect\cite{154}.
         Only statistical uncertainties are given.
         The dashed curves are $\nu^{-1}$ Regge fits to the HERMES data with 
         $W >$ 4.5 GeV; the dash-dotted curves show the next-to-leading order
         QCD parameterization \protect\cite{GRSV}.} 
\label{sigcp}
\end{minipage}
\end{center}
\end{figure}
The cross section differences $[\sigma_{1/2}-\sigma_{3/2}]$ calculated from 
the extracted values of $A_1^{\mathrm p}$ and $A_1^{\mathrm n}$  
by means of Eq.~\ref{dsigma} are presented in Fig.~\ref{sigcp}.
They are compared to the values of $\left[\sigma_{1/2}-\sigma_{3/2}\right]$ 
determined from the published data from other experiments for
$A_1^{\mathrm p}$ using polarised proton targets \cite{EMC,143,SMC} and for 
$A_1^{\mathrm n}$ using polarised helium-3 targets \cite{An,154}. 
There is good agreement between different experiments in the overlapping 
kinematic regions.
Note that the data of Refs. \cite{143,An,154} were obtained at fixed 
scattering angles and are therefore restricted to kinematic regions that are
highly correlated in $\nu$ and $Q^2$.
An evaluation of the integrals $I(Q^2)$ with these data would thus require
interpolations.

\begin{figure}[ht]
\begin{center}
   \begin{tabular}[t]{c}
\subfigure[]{\epsfxsize 4.2 cm {\epsfbox{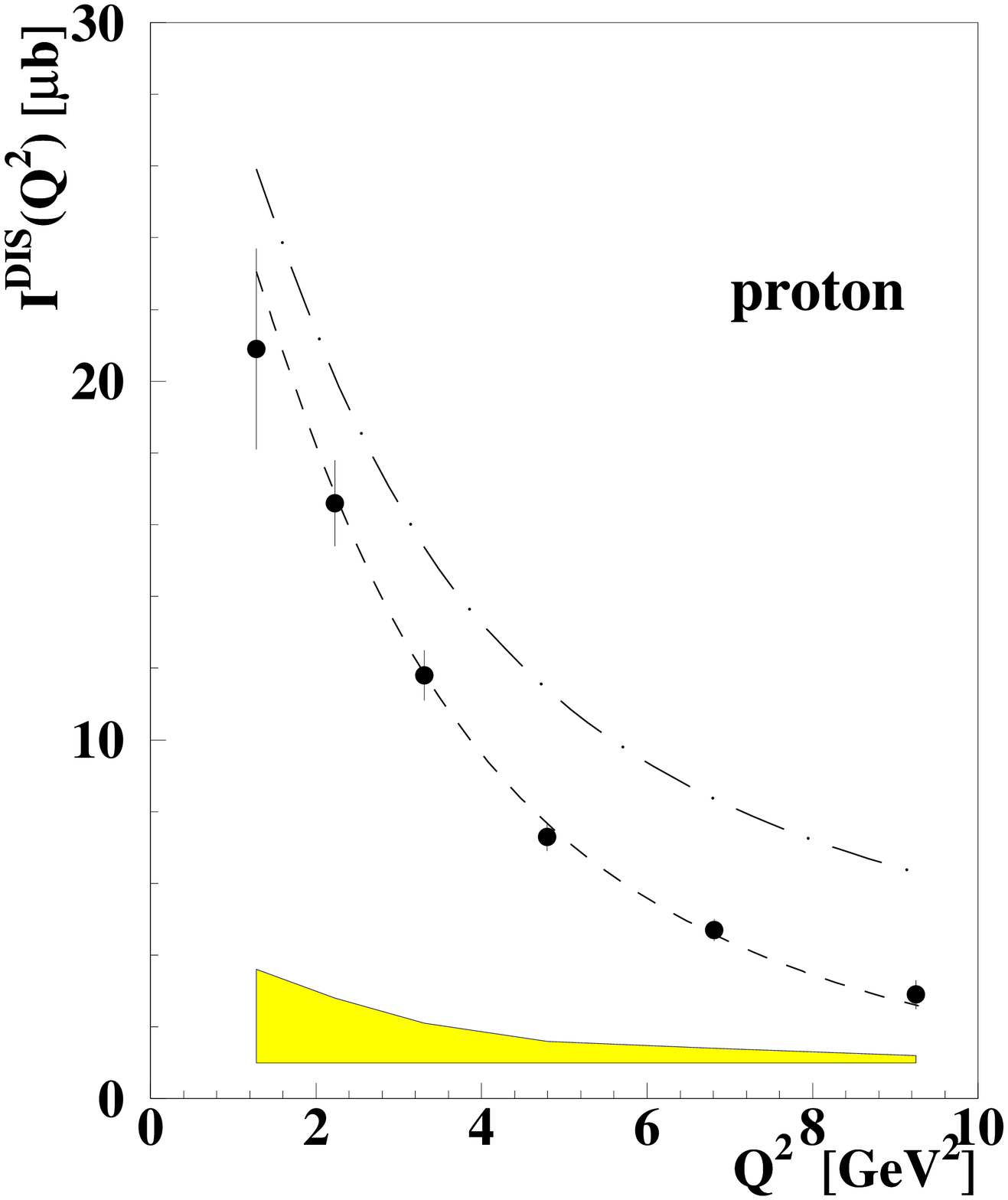}}}
\subfigure[]{\epsfxsize 4.2 cm {\epsfbox{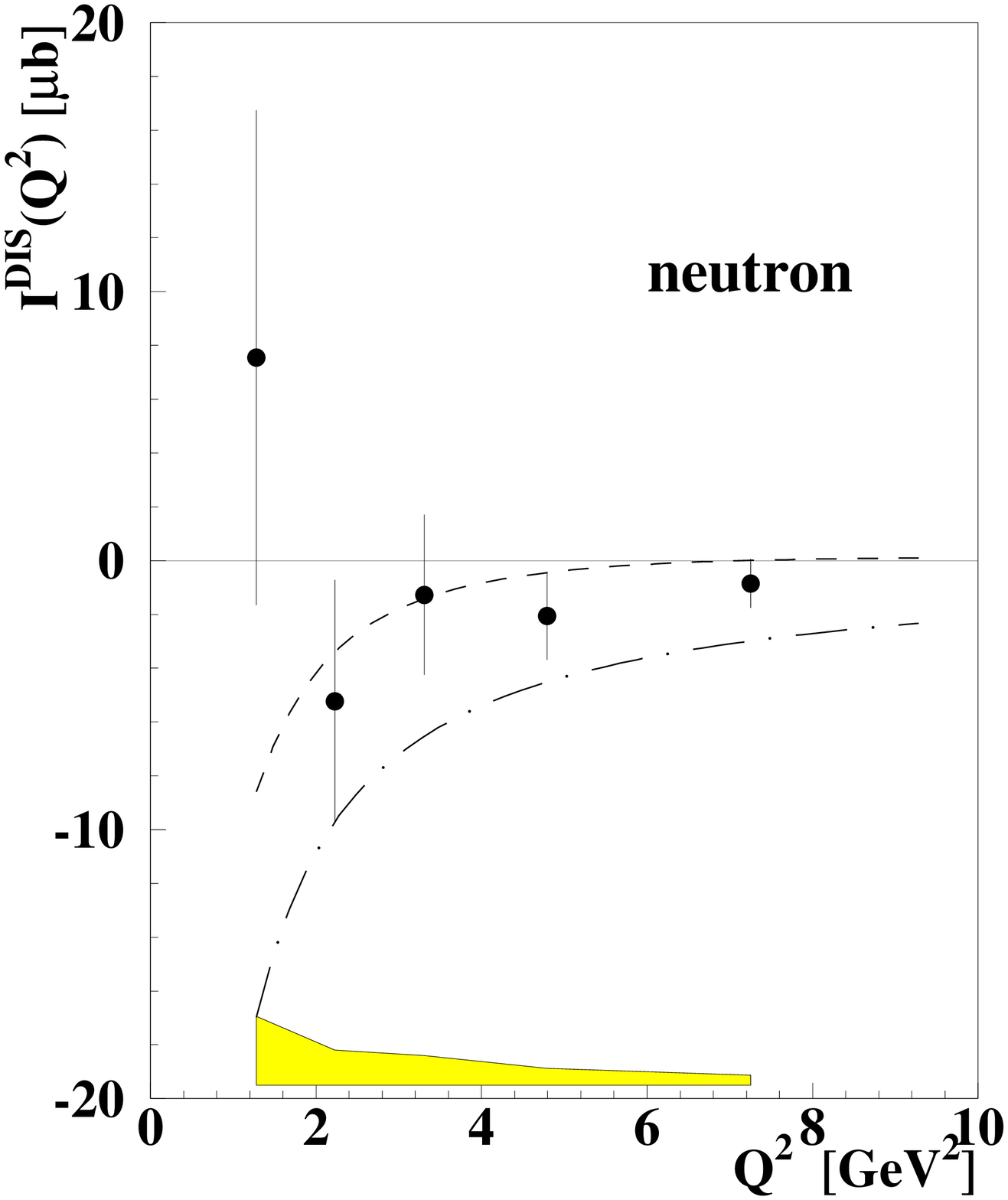}}}
   \end{tabular}
\begin{minipage}[r]{\linewidth}
\caption{The generalised GDH integral as a function of $Q^2$ in the deep
         inelastic region.
         The points are $I^{\mathrm{DIS}}_{\mathrm{HERMES}}(Q^2)$ as
         measured for HERMES data in the range $\nu_0 \leq \nu \leq$ 23.5 GeV
         for the proton (a) and for the neutron (b). 
         The error bars show the statistical uncertainties and the bands
         represent the systematic uncertainties.
         See text for the explanation of the curves.}
\label{gdhestr}
\end{minipage}
\end{center}
\end{figure}

All HERMES values for the proton are clearly positive, ranging between 
about 1 $\mu$b and about 16 $\mu$b.
Most of the data for the neutron at $\nu \geq$ 8 GeV are negative, ranging 
between about $-$10 $\mu$b and 0 $\mu$b.
In each $Q^2$-bin the present data for the cross section difference 
$[\sigma_{1/2}-\sigma_{3/2}]$ has been multiplied by $1/\nu$ 
and integrated over the range $\nu_0 < \nu <$ 23.5 GeV, 
with $\nu_0 = (W_0^2-M^2+Q^2)/2M$.
In the integration the $\nu$ dependence of the integrand $F_1/(K \nu)$ within 
the individual $\nu$-bins was fully accounted for.
The data listed in Tab.~1 and shown in Fig.~\ref{gdhestr} represent the  
DIS component for the HERMES kinematic range
$I^{\mathrm{DIS}}_{\mathrm{HERMES}}(Q^2)$ of $I(Q^2)$.
For the proton this contribution decreases from about 21 $\mu$b at
$Q^2 =$ 1.3 GeV$^2$ to about 3 $\mu$b at $Q^2 =$ 9.3 GeV$^2$. 
The contribution for the neutron is smaller in absolute value and  
between -5 $\mu$b and +8 $\mu$b.

\begin{minipage}[c]{8.0 cm}
\begin{table}[ht]
Table 1: Results on $I^{\mathrm{DIS}}_{\mathrm{HERMES}}(Q^2)$ for the proton
and neutron.
\begin{center}
\medskip
\begin{tabular}{cc}
\multicolumn{2}{c}{Proton} \\
$Q^2$ [GeV$^2$]  &   $I^{\mathrm{DIS}}_{\mathrm{HERMES}}(Q^2)$ $\pm$ stat. $\pm$ syst. [$\mu$b] \\
\hline
  1.28             & 20.9 $\pm$ 3.0 $\pm$ 2.6 \\
  2.23             & 16.6 $\pm$ 1.2 $\pm$ 1.8 \\
  3.31             & 11.8 $\pm$ 0.7 $\pm$ 1.1 \\
  4.79         &$\;$ 7.3  $\pm$ 0.4 $\pm$ 0.6 \\
  6.81         &$\;$ 4.7  $\pm$ 0.3 $\pm$ 0.4 \\
  9.25         &$\;$ 2.9  $\pm$ 0.4 $\pm$ 0.2 \\
\hline
\cline{1-2}
\multicolumn{2}{c}{Neutron} \\ 
$Q^2$ [GeV$^2$]    &        $I^{\mathrm{DIS}}_{\mathrm{HERMES}}(Q^2)$ $\pm$ stat. $\pm$ syst. [$\mu$b] \\
\hline
  1.28          &$\,$ 7.5 $\pm$ 9.2 $\pm$ 2.6 \\
  2.23             & -5.2 $\pm$ 4.5 $\pm$ 1.3 \\
  3.31             & -1.3 $\pm$ 3.0 $\pm$ 1.1 \\
  4.79             & -2.1 $\pm$ 1.6 $\pm$ 0.6 \\
  7.25             & -0.9 $\pm$ 0.9 $\pm$ 0.4 \\
\end{tabular}
\end{center}
\end{table}
\end{minipage}

The sizes of the systematic uncertainties are indicated by the bands 
at the bottom of Figs.~3a and 3b.
The main contributions are those from the beam and target polarisations. 
Other sources are uncertainties in $A_2$, in radiative 
and smearing corrections, and in the knowledge of the unpolarised structure 
functions $F_2$ and $R$. 
For the lowest $Q^2$ bin the uncertainty due to the knowledge of $A_2$
is dominant.
In addition, for the neutron there is a contribution from the nuclear 
corrections.

The data are compared with estimates of the integrals that do not
include contributions from nucleon resonances or possible higher twist 
effects.
These estimates were derived using a parameterization \cite{GRSV} for the 
asymmetry $A_1$ given by a next-to-leading order (NLO) QCD analysis.
Note that a number of the low $Q^2$, low $\nu$ data points shown in
Fig.~\ref{sigcp} from Ref.~\cite{143} are not used in this parameterisation.
The dashed curves in Fig.~\ref{gdhestr} show the integrals for this
parameterisation over the $x$ range measured at HERMES for proton and neutron.
They are in good agreement with all data, indicating that higher twist
effects do not contribute significantly in the deep inelastic region,
even at the lowest measured  $Q^2$.
The dash-dotted curves show the integrals of the NLO parameterisation from
$\nu_0$ to infinity, thus including in addition the low $x$
(high $\nu$) contribution, which for the proton is about 4$\mu$b.

An alternative approach was also used to account for the high $\nu$
region.
It is well known that the applicability of the simple Regge picture is 
uncertain and that the asymptotic behaviour of
$[\sigma_{1/2}(\nu)-\sigma_{3/2}(\nu)]$
at higher energy or $g_1(x)$ at low $x$ requires a more refined treatment
\cite{EL,QCD}.
However, if a Regge behaviour is assumed here for the sake of comparison,
the main high $\nu$ contribution comes from the $a_1$(1260) Regge trajectory
such that $[\sigma_{1/2}-\sigma_{3/2}] \propto \nu^{\alpha-1}$, where $\alpha$
is the intercept of this trajectory.
The present data with $W \geq 4.5$ GeV have been fitted with such a 
parameterisation using $\alpha$= 0 \cite{Heim}.
Fits are shown as dashed curves in Fig.~\ref{sigcp}; they are in fair 
agreement with data.
The contribution of the high $\nu$ Regge extrapolation to 
the integral is about 3 $\mu$b for the proton data, as well as for the 
neutron data with $Q^2 <$ 4 GeV$^2$.

Little experimental information is available about the size of the
resonance contribution to $I(Q^2)$. 
Up to now the resonance part ($W<$ 2 GeV)  of $\Gamma_1$ has been
determined only for two $Q^2$ values (0.5 GeV$^2$ and 1.2 GeV$^2$)
\cite{143,Kuhn}. 
This contribution has been compared to the part of $\Gamma_1$ 
at larger $W$ and at the same fixed $Q^2$, obtained by interpolating 
to the appropriate $Q^2$ the data taken at fixed scattering angles of 
4.5 and 7 degrees and at the two beam energies 9.7 GeV and 16.2 GeV. 
For example at $Q^2$ = 1.2 GeV$^2$, the resonance contribution was found 
to be about 37\% of the whole integral $\Gamma_1$ for the proton, including
a negative contribution from the first resonance \cite{143,Baum}.

A measurement of $A_1$ and $A_2$ is planned at TJNAF \cite{EX97} for a 
determination of the contribution of the resonance region to the generalised 
GDH integral $I(Q^2)$ for $Q^2<$ 3 GeV$^2$. 
This together with the present results in the deep inelastic scattering regime 
and a more refined estimate of the high energy extrapolation will provide
the Q$^2$ evolution of the entire integral.
In addition, several experiments are planned at different facilities 
to measure the spin-dependent photo-production cross sections 
to test the GDH sum rule for real photons \cite{bia}.

In conclusion, the polarised cross section differences 
$[\sigma_{1/2}-\sigma_{3/2}]$ have been determined for the proton and neutron
in the kinematic range 0.8 GeV$^2$ $< Q^2 <$ 12 GeV$^2$, $W >$ 1.8 GeV, 
$\nu <$ 23.5 GeV, and the corresponding DIS parts of the generalised GDH 
integrals $I(Q^2)$ have been evaluated. 
For $Q^2$ above 1 GeV$^2$ the results are at least of the same order of 
magnitude as expectations for the resonance part of $I(Q^2)$.
Therefore the rapid excursion of $I(Q^2)$ towards the predicted negative 
values at $Q^2$=0 must occur below a $Q^2$ of about 1 GeV$^2$.
Furthermore the data are consistent with a QCD next-to-leading order fit
based on previous deep inelastic scattering data, indicating that higher
twist effects are not significant even at the lowest measured $Q^2$.

We gratefully acknowledge the DESY management for its support and the 
staffs at DESY and the collaborating institutions for their significant 
effort. 
This work was supported by the Fund for Scientific Research-Flanders (FWO) 
of Belgium; the Natural Sciences and Engineering Research Council of Canada;
the INTAS, HCM and TMR contributions from the European Community;
the German Bundesministerium f\"{u}r Bildung, Wissenschaft, Forschung 
und Technologie (BMBF), the Deutscher Akademischer Austauschdienst (DAAD); 
the Italian Istituto Nazionale di Fisica Nucleare (INFN);
Monbusho, JSPS, and Toray Science Foundation of Japan;
the Dutch Stichting voor Fundamenteel Onderzoek der Materie (FOM);
the UK Particle Physics and Astronomy Research Council; and
the US Department of Energy and National Science Foundation. 

\vskip 1cm
\noindent
$^\dagger$ Deceased.

\end{multicols}

\begin{thebibliography}{}

\bibitem{Geras} 
S.B.~Gerasimov, Sov. J. Nucl. Phys. {\bf 2} (1966) 430.
\bibitem{DreHea} 
S.D.~Drell and A.C.~Hearn, Phys. Rev. Lett. {\bf 16} (1966) 908.
\bibitem{GGT}
M.~Gell-Mann {\em et al.}, Phys. Rev. {\bf 95} (1954) 1612.
\bibitem{LOW}
F.E.~Low, Phys. Rev. {\bf 96} (1954) 1428.
\bibitem{GIL}
F.J.~Gilman, Phys. Rev. {\bf 167} (1968) 1365.
\bibitem{ANSELM}
M. ~Anselmino, B.L.~Ioffe and E.~Leader, Sov. J. Nucl. Phys. {\bf 49} (1989) 
136.
\bibitem{EMC}
EMC Collaboration,
J.~Ashman {\em et al.}, Nucl. Phys. {\bf B 328} (1989) 1.
\bibitem{E130}
E130 Collaboration, G.~Baum {\em et al.}, Phys. Rev. Lett. {\bf 51} (1983) 
1135.
\bibitem{EJ}
J.~Ellis and R.L.~Jaffe, Phys. Rev. {\bf D 9} (1974) 1444;
Phys. Rev. {\bf D 10} (1974) 1669.
\bibitem{B1}
V.D.~Burkert and B.L.~Ioffe, Phys. Lett. {\bf B 296} (1992) 223.
\bibitem{B2}
V.D.~Burkert and Z.~Li, Phys. Rev. {\bf D 47} (1993) 46.
\bibitem{LI2}
Z.~Li, Phys. Rev. {\bf D 47} (1993) 1854.
\bibitem{B3}
J.~Soffer and O.~Teryaev, Phys. Rev. Lett. {\bf 70} (1993) 3373.
\bibitem{B4}
X.~Ji, Phys. Lett. {\bf B 309} (1993) 187.
\bibitem{LI}
Z.~Li {\em et al.}, Phys. Rev. {\bf D 46} (1992) 70.
\bibitem{LILI}
Z.~Li and Z.~Li, Phys. Rev. {\bf D 50} (1994) 3119.
\bibitem{HERMES} 
HERMES Collaboration,
K.~Ackerstaff {\em et al.}, Phys. Lett. {\bf B 404} (1997) 383.
\bibitem{A1p}
HERMES Collaboration,
A.~Airapetian {\em et al.}, DESY 98-072 (1998), hep-ex 9807015 (1998),
Phys. Lett. {\bf B} (in press).
\bibitem{ST}
A.A.~Sokolov and I.M.~Ternov, Sov. Phys. Doklady {\bf 8} (1964) 1203.
\bibitem{BA}
D.P.~Barber {\em et al.}, Phys. Lett. {\bf B 343} (1997) 436.
\bibitem{STO}
F.~Stock {\em et al.}, Nucl. Instr. and Meth. {\bf 343} (1994) 334.
\bibitem{xx97}
D.~De Schepper {\em et al.},  MIT-LNS preprint 01/97, (1997).
\bibitem{yy97}
HERMES Collaboration,
K.~Ackerstaff {\em et al.}, DESY 98-057 (1998), hep-ex 9806008 (1998),
Nucl. Instr. and Meth. {\bf A} (in press).
\bibitem{NMCF2} 
NMC Collaboration,
M.~Arneodo {\em et al.}, Phys. Lett. {\bf B 364} (1995) 107.
\bibitem{WHITL}
L.W.~Whitlow {\em et al.},  Phys. Lett. {\bf B 250} (1990) 193.
\bibitem{A2}
E154 Collaboration,
K.~Abe {\em et al.}, Phys. Lett. {\bf B 404} (1997) 377.
\bibitem{143}
E143 Collaboration,
K.~Abe {\em et al.}, Phys. Rev. Lett. {\bf 76} (1996) 587;
K.~Abe {\em et al.}, SLAC-PUB-7753, hep-ph 9802357 (1998).
\bibitem{SMC}
SMC Collaboration,
B.~Adeva {\em et al.}, Phys. Lett. {\bf B 412} (1997) 414;
CERN-EP 98/085 (1998). 
\bibitem{POLRAD}
I.V.~Akushevich and N.M.~Shumeiko, J. Phys. {\bf G 20} (1994) 513.
\bibitem{FRIAR}
J.L.~Friar {\em et al.}, Phys. Rev. {\bf C 42} (1990) 2310;
C.~Ciofi degli Atti {\em et al.}, Phys. Rev. {\bf C 48} (1993) R968.
\bibitem{E143Q2}
E143 Collaboration,
K.~Abe {\em et al.}, Phys. Lett. {\bf B 364} (1995) 61.
\bibitem{An}
E142 Collaboration,
P.L.~Anthony {\em et al.}, Phys. Rev. {\bf D 54} (1996) 6620. 
\bibitem{154}
E154 Collaboration,
K.~Abe {\em et al.}, Phys. Rev. Lett. {\bf 79} (1997) 26.
\bibitem{GRSV}
M.~Gl\"uck {\em et al.}, Phys. Rev. {\bf D 53} (1996) 4775.
\bibitem{EL}
E.~Leader {\em et al.}, hep-ph 9807251 (1998).
\bibitem{QCD}
E154 Collaboration,
K.~Abe {\em et al.}, Phys. Lett. {\bf B 405} (1997) 180;
SMC Collaboration CERN-EP 98-086 (1998).
\bibitem{Heim}
R.L.~Heimann, Nucl. Phys. {\bf B 64} (1973) 429.
\bibitem{Kuhn}
E143 Collaboration,
K.~Abe {\em et al.}, Phys. Rev. Lett. {\bf 78} (1997) 815.
\bibitem{Baum}
G.~Baum {\em et al.}, Phys. Rev. Lett. {\bf 45} (1980) 2000.
\bibitem{EX97}
V.D.~Burkert {\em et al.}, CEBAF PR-91-23 (1991); 
S.~Kuhn {\em et al.}, CEBAF PR-93-09 (1993); 
Z.E.~Meziani {\em et al.}, CEBAF PR-94-10 (1994);
J.P.~Chen {\em et al.}, TJNAF PR-97-110 (1997).
\bibitem{bia}
N.~Bianchi, Proc. Workshop on Electron Nucleus Scattering,
O.~Benhar and A.~Fabrocini Eds., Elba, July 1-5, 1996, p.363 and 
references therein.
 
\end{thebibliography}
\end{document}